\documentclass[letterpaper,aps,pre,reprint,superscriptaddress,amsmath,showpacs]{revtex4-1}

\usepackage[hyperindex,breaklinks,colorlinks=true]{hyperref}
\usepackage{graphicx}

\begin{document}

\title{Comment on ``Fluctuation theorem for hidden entropy production''}

\author{Yongjoo Baek}
\affiliation {Department of Physics,
Korea Advanced Institute of Science and Technology, Daejeon 305-701, Korea}

\author{Meesoon Ha}
\affiliation {Department of Physics Education,
Chosun University, Gwangju 501-759, Korea}

\author{Hawoong Jeong}
\affiliation{Department of Physics and Institute for the BioCentury,
Korea Advanced Institute of Science and Technology, Daejeon 305-701, Korea}
\affiliation{APCTP, Pohang, Gyeongbuk 790-784, Korea}

\author{Hyunggyu Park}
\affiliation{School of Physics,
Korea Institute for Advanced Study, Seoul 130-722, Korea}

\date{\today}

\begin{abstract}
Recently, Kawaguchi and Nakayama (KN) [\href{http://link.aps.org/doi/10.1103/PhysRevE.88.022147}{Phys. Rev. E {\bf 88}, 022147 (2013)}] showed that the hidden entropy production associated with a coarse-graining procedure obeys the integral fluctuation theorem (IFT) if the original process does not involve any odd-parity variable that changes its sign under time reversal. This was interpreted as the evidence that odd-parity variables play an important role in the derivation of irreversible stochastic dynamics from time-reversible deterministic dynamics. In this Comment, we show that KN's approach is inadequate for describing the origin of irreversible stochastic dynamics, which calls into question whether odd-parity variables are required for the emergence of macroscopic irreversibility.
\end{abstract}

\pacs{05.20.-y, 05.70.-a}

\maketitle

Kawaguchi and Nakayama (KN)~\cite{Kawaguchi2013} recently claimed that the hidden entropy production, or the amount of entropy production ignored by coarse-graining, obeys the integral fluctuation theorem (IFT) when the original process involves only even-parity variables, which are time-reversal invariant. They also claimed that the IFT is not generally valid in the presence of odd-parity variables, which are antisymmetric under time reversal. Specifically, the density function asymmetry for odd-parity variables was pointed out as the key source of the IFT violation, which is necessary for the increased mean entropy production after coarse-graining. Based on these claims, KN concluded that odd-parity variables play an important role in the emergence of irreversible stochastic dynamics from time-reversible Hamiltonian dynamics.

In this Comment, we point out that (1) KN's coarse-grained entropy production does not properly indicate the irreversibility of the coarse-grained process and that (2) KN's coarse-graining scheme neglects the memory cutoff, which is necessary for deriving a stochastic description of a deterministic system. These problems suggest that KN's IFT for hidden entropy production is not relevant to the origin of irreversible stochastic dynamics.

We start with a brief review of KN's approach. Consider a Markov process of state variables $x$ and $y$ controlled by a time-dependent protocol $\lambda(t)$. Here each of $x$ and $y$ may represent multiple variables. The process starts at $t = 0$ and ends at $t = \tau$. For convenience, we define a path $\{\mathbf{x},\mathbf{y}\}$ as $\{\mathbf{x}(t),\mathbf{y}(t)\} \equiv \{(x_t, y_t): t \in [0,\tau]\}$, whose initial and final points are distributed by $P_0(x_0,y_0)$ and $P_\tau(x_\tau,y_\tau)$, respectively. Then, the probability of a given path $\{\mathbf{x},\mathbf{y}\}$ can be written as
\begin{equation}
P_{\lambda} [\mathbf{x}, \mathbf{y}]
= P_0 (x_0, y_0) W_{\lambda} [\mathbf{x}, \mathbf{y} | x_0, y_0],
\label{eq:p-for}
\end{equation}
where $W_{\lambda} [\mathbf{x}, \mathbf{y} | x_0, y_0]$ is the conditional probability for the path $\{\mathbf{x},\mathbf{y}\}$ starting from $(x_0,y_0)$. In a Markov process, this conditional path probability can be factorized into an infinite product of infinitesimal transition probabilities.

To characterize the irreversibility of the path $\{\mathbf{x},\mathbf{y}\}$, we define the corresponding time-reverse path $\{\mathbf{x}^\dagger,\mathbf{y}^\dagger\}$ as $\mathbf{x}^\dagger(t) = \bar{\mathbf{x}}(\tau-t)$, where $\bar{\mathbf{x}}$ represents the mirrored path with an extra minus sign for each odd-parity variable~\cite{Spinney2012a,*Spinney2012b,*Ford2012,Lee2013}. The time-reverse path starts at the mirror state of the end point of the original path, $(x_0^\dagger, y_0^\dagger)=(\bar{x}_\tau, \bar{y}_\tau)$, whose distribution is denoted as $P_0^\dagger (x_0^\dagger, y_0^\dagger)$. Then, the probability for the time-reverse path is given as
\begin{equation}
P_{\lambda^\dagger} [\mathbf{x}^\dagger, \mathbf{y}^\dagger]
= P_0^\dagger (x_0^\dagger, y_0^\dagger) W_{\lambda^\dagger} [\mathbf{x}^\dagger, \mathbf{y}^\dagger | x_0^\dagger, y_0^\dagger],
\label{eq:p-rev}
\end{equation}
where $\lambda^\dagger(t)=\lambda(\tau-t)$ for the proper time-reverse process.

Since the process is Markovian, the total entropy production along a path $\{\mathbf{x},\mathbf{y}\}$ is given by the formula~\cite{Seifert2005}
\begin{equation}
\Sigma(\mathbf{x}, \mathbf{y}) \equiv \ln \frac{P_0(x_0,y_0)}{P_\tau(x_\tau,y_\tau)} + \ln \frac{W_{\lambda} [\mathbf{x}, \mathbf{y} | x_0, y_0]}{W_{\lambda^\dagger} [\mathbf{x}^\dagger, \mathbf{y}^\dagger | x_0^\dagger, y_0^\dagger]},
\label{eq:sig}
\end{equation}
where the first term is the Shannon entropy change of the system, and the second term is the entropy production of the environment which is assumed to equilibrate instantaneously~\cite{Hinrichsen2011}.

KN defined coarse-graining as integration over a subset of state variables. This means that the probabilities of coarse-grained paths are given as
\begin{align}
\tilde{P}_{\lambda} [\mathbf{x}] &\equiv \int \mathrm{d}\mathbf{y} \,P_\lambda [\mathbf{x}, \mathbf{y}], \label{eq:cg-for} \\
\tilde{P}_{\lambda^\dagger} [\mathbf{x}^\dagger] &\equiv \int \mathrm{d}\mathbf{y}^\dagger\,P_{\lambda^\dagger} [\mathbf{x}^\dagger, \mathbf{y}^\dagger]. \label{eq:cg-rev}
\end{align}
The coarse-grained state distributions $\tilde{P}_0$, $\tilde{P}_\tau$, and $\tilde{P}_0^\dagger$ are similarly defined as the marginal distributions of $P_0$, $P_\tau$, and $P_0^\dagger$, respectively. Then, the conditional probabilities for coarse-grained paths can be written as
\begin{align}
\tilde{W}_{\lambda} [\mathbf{x} | x_0, P_0(x_0,y_0)] &\equiv \tilde{P}_{\lambda} [\mathbf{x}]/\tilde{P}_0(x_0), \label{eq:cg-con-for}\\
\tilde{W}_{\lambda^\dagger} [\mathbf{x}^\dagger | x_0^\dagger, P_0^\dagger(x_0^\dagger,y_0^\dagger)] &\equiv \tilde{P}_{\lambda^\dagger} [\mathbf{x}^\dagger]/\tilde{P}_0^\dagger(x_0^\dagger), \label{eq:cg-con-rev}
\end{align}
where $P_0$ and $P_0^\dagger$ are respectively included in the arguments of $\tilde{W}_\lambda$ and $\tilde{W}_{\lambda^\dagger}$ to indicate the dependence of the latter on the former.

In the manner of Eq.~(\ref{eq:sig}), KN proposed that the coarse-grained entropy production can be similarly defined as
\begin{equation} \label{eq:cg-sig}
\tilde\Sigma[\mathbf{x}] \equiv \ln \frac{\tilde{P}_0(x_0)}{\tilde{P}_\tau(x_\tau)} + \ln \frac{\tilde{W}_{\lambda} [\mathbf{x} | x_0, P_0(x_0,y_0)]}{\tilde{W}_{\lambda^\dagger} [\mathbf{x}^\dagger | x_0^\dagger, P_0^\dagger(x_0^\dagger,y_0^\dagger)]}.
\end{equation}
As pointed out by KN, the coarse-grained process obtained from Eqs.~(\ref{eq:cg-for}) and (\ref{eq:cg-rev}) is generally non-Markovian, so the above definition may not satisfy some properties of the conventional entropy production, such as the additivity over time. Nevertheless, KN adopted Eq.~(\ref{eq:cg-sig}) as a natural extension of Eq.~(\ref{eq:sig}) to non-Markovian processes.

If we accept Eq.~(\ref{eq:cg-sig}), the hidden entropy production can now be defined as
\begin{equation} \label{eq:xi}
\Xi [\mathbf{x}, \mathbf{y}] \equiv \Sigma[\mathbf{x}, \mathbf{y}] - \tilde{\Sigma}[\mathbf{x}].
\end{equation}
Using Eqs. (\ref{eq:p-for}) -- (\ref{eq:xi}), we obtain
\begin{align} \label{eq:ift}
\langle e^{-\Xi}\rangle
&= \int \mathrm{d}\mathbf{x}\,\mathrm{d}\mathbf{y}\, P_\lambda [\mathbf{x}, \mathbf{y}] \nonumber \\
&\quad \times \frac{P_\tau (x_\tau,y_\tau) W_{\lambda^\dagger} [\mathbf{x}^\dagger, \mathbf{y}^\dagger|x_0^\dagger,y_0^\dagger]}{P_0(x_0,y_0)W_\lambda [\mathbf{x}, \mathbf{y}|x_0,y_0]} \nonumber \\
&\quad \times \frac{\tilde{P}_0(x_0)\tilde{W}_{\lambda} [\mathbf{x}|x_0,P_0(x_0,y_0)]}{\tilde{P}_{\tau}(x_\tau)\tilde{W}_{\lambda^\dagger} [\mathbf{x}^\dagger|x_0^\dagger,P_0^\dagger(x_0^\dagger,y_0^\dagger)]} \nonumber \\
&= \int \mathrm{d}\mathbf{x} \, \frac{\tilde{W}_{\lambda^\dagger}[\mathbf{x}^\dagger|x_0^\dagger,P_\tau(x_\tau,y_\tau)]}{\tilde{W}_{\lambda^\dagger}[\mathbf{x}^\dagger|x_0^\dagger,P_0^\dagger(x_0^\dagger,y_0^\dagger)]} \tilde{P}_\lambda[\mathbf{x}],
\end{align}
where $\tilde{W}_{\lambda^\dagger}[\mathbf{x}^\dagger|x_0^\dagger,P_\tau(x_\tau,y_\tau)]$ is the conditional probability of a coarse-grained time-reverse path $\mathbf{x}^\dagger$ starting from $x_0^\dagger$ distributed by $P_0^\dagger (x_0^\dagger, y_0^\dagger) = P_\tau (x_\tau, y_\tau)$. Hence, the IFT $\langle e^{-\Xi}\rangle = 1$ is satisfied if
\begin{equation} \label{eq:ift-con}
P_0^\dagger ({x}_0^\dagger,{y}_0^\dagger) = P_{\tau} (x_\tau, y_\tau),
\end{equation}
which indicates that the definition of $P_0^\dagger$ determines the sufficient conditions for the IFT.

We note that KN defined $P_0^\dagger$ as~\cite{exp1}
\begin{equation} \label{eq:ini-con-kn}
P_0^\dagger ({x}_0^\dagger,{y}_0^\dagger) \equiv P_{\tau} (\bar{x}_\tau, \bar{y}_\tau),
\end{equation}
which means that the time-reverse process is a continuation of the forward process under the time-reverse protocol. This definition ensures that $\tilde{\Sigma}$ is empirically observable, because the initial state of the time-reverse process can be simply prepared by the forward process. The IFT is satisfied when Eqs.~(\ref{eq:ift-con}) and (\ref{eq:ini-con-kn}) are consistent, so the mirror symmetry $P_{\tau}(\bar{x}_\tau,\bar{y}_\tau) = P_{\tau}(x_\tau,y_\tau)$ becomes a sufficient condition for the IFT. Since $\langle \tilde{\Sigma} \rangle > \langle \Sigma \rangle$ requires the violation of the IFT, KN claimed that the absence of the mirror symmetry is crucial for the emergence of irreversible stochastic dynamics from time-reversible deterministic dynamics.

KN's claim relies on the premises that $\tilde{\Sigma}$ represents the irreversibility of the coarse-grained process and that the coarse-graining procedure defined by Eqs.~(\ref{eq:cg-for}) and (\ref{eq:cg-rev}) can convert a deterministic process to a stochastic one. However, as we discuss below, both premises are not true.

First, $\tilde{\Sigma}$ is not a good indicator of irreversibility as long as $P_0^\dagger$ is defined by Eq.~(\ref{eq:ini-con-kn}). To check the reversibility of a coarse-grained path reaching the final state $x_\tau$, the ensemble of coarse-grained time-reverse paths must contain those starting from $x_0^\dagger \equiv \bar{x}_\tau$. Otherwise, it is not even possible to quantify the relative difficulty of the time-reverse path with respect to the forward path, which is mathematically reflected in the fact that $\tilde{W}_{\lambda^\dagger}$ (and therefore $\tilde{\Sigma}$ as well) is ill defined if both $\tilde{P}_0(x_0^\dagger)$ and $\tilde{P}_{\lambda^\dagger}(\mathbf{x}^\dagger)$ are strictly zero. Such cases may indeed occur if Eq.~(\ref{eq:ini-con-kn}) is used, because $\tilde{P}_\tau(x_\tau) > 0$ does not always imply $\tilde{P}_0(x_0^\dagger) = \tilde{P}_\tau(\bar{x}_\tau) > 0$. To address this problem, $P_0^\dagger$ must be redefined to satisfy
\begin{equation} \label{eq:ini-con-mod}
\tilde{P}_0^\dagger (x_0^\dagger) = \tilde{P}_{\tau} (x_\tau),
\end{equation}
which automatically ensures that $\tilde{\Sigma}(\mathbf{x})$ is well defined for every coarse-grained path $\mathbf{x}$ of nonzero probability. This definition of $\tilde{P}_0^\dagger$ implies that the time-reverse process is initiated by time-reversing all variables of interest (represented by $x$) at the end of the forward process, which is consistent with the requirement discussed in \cite{Spinney2012c}.

Second, the coarse-graining scheme defined by Eqs.~(\ref{eq:cg-for}) and (\ref{eq:cg-rev}) cannot convert a deterministic process to a stochastic one. This can be easily checked by considering a deterministic process, whose path is uniquely determined by its initial state as $\{\mathbf{\hat{x}}(x_0,y_0),\mathbf{\hat{y}}(x_0,y_0)\}$, so that the conditional path probability can be written as
\begin{equation}
W_{\lambda} [\mathbf{x}, \mathbf{y} | x_0, y_0] = \delta [\mathbf{x}-\mathbf{\hat{x}}(x_0,y_0)]\cdot\delta [\mathbf{y}-\mathbf{\hat{y}}(x_0,y_0)],
\end{equation}
where $\delta$ represents an infinite product of delta functions over the path $\{\mathbf{{x}},\mathbf{{y}}\}$. Using Eqs.~(\ref{eq:p-for}) and (\ref{eq:cg-for}), the probability of a coarse-grained path is given by
\begin{equation} \label{eq:det_for}
\tilde{P}_{\lambda} [\mathbf{x}] = \int \mathrm{d}{y}_0 \, P_0 (x_0, y_0)~
\delta [\mathbf{x}-\mathbf{\hat{x}}(x_0,y_0)].
\end{equation}
This coarse-grained process is still deterministic except for the uncertainty associated with the initial state variable $y_0$. Once the initial uncertainty is resolved by fixing $y_0$, the everlasting memory of the initial state ensures that the rest of the coarse-grained process has no random elements. Thus, despite KN's claim that the violation of the IFT for $\Xi$ allows the emergence of a random walk from a deterministic jumping process, the coarse-graining scheme involved in the definition of $\Xi$ is unable to derive the former from the latter. In general, in order to derive a stochastic equation of motion (such as the Langevin equation) from Hamiltonian dynamics, the contribution from the initial state of the hidden variables must be approximated as random noise~\cite{Zwanzig2001}. This approximation requires that the memory of the initial state is truncated at some point. Without implementing such memory cutoff, KN's IFT for $\Xi$ is irrelevant to the origin of stochasticity.

These two problems suggest that KN's IFT for $\Xi$ does not justify the attribution of macroscopic irreversibility to odd-parity variables. However, we should also note that these problems do not prove the irrelevance of odd-parity variables either. They only show that there may be other factors contributing to the increased entropy production after coarse-graining, such as the additional randomness arising from the memory cutoff. It is telling that the Evans--Searles fluctuation theorem, which provides a different explanation for the macroscopic irreversibility, does not require the existence of odd-parity variables~\cite{Sevick2008}. This may suggest that the factors neglected by KN are indeed more crucial.

In conclusion, KN's IFT for the hidden entropy production does not answer the question of how irreversible stochastic dynamics can be deduced from time-reversible deterministic dynamics. To address this question, the initial condition for the time-reverse process and the coarse-graining scheme has to be redefined, so that irreversibility and stochasticity of the coarse-grained process is properly represented in the theoretical framework.

This work was supported by the NRF Grant No.~2011-0028908(Y.B.,H.J.), 2011-0011550(M.H.), and 2013R1A1A2A10009722(H.P.).

\bibliography{EPK1014}

\end{document}